\documentclass{article}
\usepackage{authblk}
\usepackage{amsmath}
\usepackage{url}
\usepackage{graphicx}
\usepackage[utf8]{inputenc}

\usepackage{comment} 
\usepackage{amssymb} 
\usepackage{xcolor} 
\newtheorem{proposition}{Proposition}

\title{Pearson Correlations on Networks:\\Corrigendum}
\author[1]{Michele Coscia}
\author[2]{Karel Devriendt}

\affil[1]{CS Department, IT University of Copenhagen, Rued Langgaards Vej 7, Copenhagen, DK}
\affil[2]{Max Planck Institute for Mathematics in the Sciences, Inselstraße 22, Leipzig, DE}

\date{}

\begin{document}

\maketitle

\begin{abstract}
Recently, the first author proposed a measure to calculate Pearson correlations for node values expressed in a network, by taking into account distances or metrics defined on the network. In this technical note, we show that using an arbitrary choice of distances might result in imaginary or unbounded correlation values, which is undesired. We prove that this problem is solved by restricting to a special class of distances: negative type metrics. We also discuss two natural classes of negative type metrics on graphs, for which the network correlations are properly defined.
\end{abstract}

\section{Introduction}
In \cite{coscia2021pearson} the first author extended the familiar Pearson linear correlation coefficient to the case in which the correlated vectors live on a graph. In practice, this means that each entry in the vector corresponds to a node in the network. Then, to estimate the correlation, one compares the value for a node in a vector with the values of all other nodes in the network from the other vector, proportionally to their network distance. This is opposed to the regular Pearson correlation, where only the values of the corresponding entries are compared with each other, without checking their neighbors.

To compute the extended Pearson correlation, one needs to specify a weight matrix $W$, which reflects the structure of the network. The original paper fails to discuss properly the criteria one should follow when picking $W$. This is important, because it is not just a matter of performance or intuitiveness: some values of $W$ are invalid.

In this technical note we show that there is a strong constraint on $W$. Specifically, $W$ must be positive definite for vectors with zero sum, otherwise it would generate imaginary or infinite network Pearson correlation values. These problems can arise for the $W$ suggested as default in \cite{coscia2021pearson}, which is a weight matrix defined based on the shortest path distances in the graph.

We also show that this constraint is not insurmountable: all we need to obtain a proper $W$ is to use a negative type metric instead of the shortest-path distance. There are a few intuitive options to create positive definite $W$s. We propose two examples, one based on effective resistances \cite{klein1993resistance, devriendt2022effective}, the other on Euclidean node embeddings \cite{perozzi2014deepwalk, coscia2021atlas}.

The code and data to verify the claims of this note are publicly available at \url{http://www.michelecoscia.com/?page_id=2268}.

\section{Network Correlation}
The classical Pearson linear correlation coefficient is defined as:

\begin{equation} \label{eq:rho}
\rho_{x,y} = \dfrac{\sum I \times (\hat{x} \otimes \hat{y})}{\sigma_x \sigma_y},
\end{equation}

where:

\begin{itemize}
\item $I$ is the identity matrix;
\item $\hat{x} = x - \bar{x}$ is the centered version of $x$, i.e. $x$ minus its mean $\bar{x}$;
\item $\times$ is the elementwise matrix product;
\item $\otimes$ is the outer product;
\item $\sigma_x = \sqrt{\sum I \times (x \otimes x)}$ is the standard deviation of $x$.
\end{itemize}

We use this notation because it makes it easier to extend it to the network case. In practice, multiplying $I$ with $\hat{x} \otimes \hat{y}$ results in considering only its main diagonal, whose sum is the covariance of $x$ and $y$.

Now suppose that we have a graph $G = (V, E)$ and that both $x$ and $y$ are vectors of length $|V|$ -- i.e. they record one value per node of $G$. We can extend Equation \ref{eq:rho} to take into account the topology of $G$. Specifically, a valid measure of network co-variance will weight each entry $ij$ of $\hat{x} \otimes \hat{y}$ proportionally to some function of the network distance between nodes $i$ and $j$ in $G$, rather than only considering the diagonal $ii$ entries like in the regular Pearson correlation.

We can do so by specifying a proper matrix $W$, and define the network Pearson correlation as:

\begin{equation} \label{eq:rho-net}
\rho_{x,y,G} = \dfrac{\sum W \times (\hat{x} \otimes \hat{y})}{\sigma_{x, W} \sigma_{y, W}},
\end{equation}

with $\sigma_{x,W} = \sqrt{\sum W \times (x \otimes x)}$.

The original paper  \cite{coscia2021pearson} proposes $W = e^{-kP}$, where $P_{ij}$ is the shortest path length between nodes $i$ and $j$ in $G$ and where the exponent is taken entrywise. $k$ is a simple scaling factor, with default value $k = 1$ -- when it applies, we omit $k$ for simplicity as it does not affect any of the arguments in this note.

\section{Problem of Negative Variances}
Let us consider the formulation of $\sigma_{x,W}$. This is the square root of the network variance of $x$ and it appears in the denominator of the network Pearson correlation. It follows that $\sum W \times (x \otimes x)$ must be a strictly positive value: if it were negative, Equation \ref{eq:rho-net} evaluates as an imaginary number, which would be hardly interpretable as a network correlation and, if it were zero, Equation \ref{eq:rho-net} would become infinity.

\subsection{Conditions on the Weight Matrix $W$}
The above observations lead to the following conditions on $W$ that make the network correlation well-defined.
\begin{proposition}\label{prop:definite}
The network Pearson correlation \eqref{eq:rho-net} is well-defined (real and finite) for all non-constant input vectors $x,y$ if and only if the weight matrix $W$ is positive definite on $\operatorname{span}(1)^\perp$. Furthermore, in this case the network Pearson correlation is a real number in $[-1,1]$.
\end{proposition}
\textbf{Proof}
By ``well-defined" we mean that the network Pearson correlation $\rho_{x,y,G}$ is real and finite for all real input vectors $x,y$ that are not constant.

First, the numerator of the network Pearson correlation $\rho_{x,y,G}$ is a quadratic form over the reals determined by a real, symmetric matrix $W$. This produces a finite real number as necessary.

Second, the denominator of the network Pearson correlation equals 
\begin{equation}\label{eq: denominator}
\sigma_{x,W}\sigma_{y,W} = \sqrt{\left(\sum W_{ij}\hat{x}_i\hat{x}_j\right)\left(\sum W_{ij}\hat{y}_i\hat{y}_j\right)},
\end{equation}
i.e. the product of two quadratic forms determined by the matrix $W$. We note that the normalization $x\rightarrow \hat{x}=x-\bar{x}$ leads to $\sum \hat{x}=0\Rightarrow \hat{x}\perp 1$, which means that the quadratic forms are defined on $\operatorname{span}(1)^\perp$. For $\rho_{x,y,G}$ to be well-defined, both quadratic forms in \eqref{eq: denominator} need to be nonnegative -- if one were negative, then the square root would yield an imaginary number -- and nonzero -- if one were zero then this would yield a division by zero; in other words, the quadratic form needs to be \textit{strictly positive} on $\operatorname{span}(1)^\perp$, which is equivalent to $W$ being positive definite on $\operatorname{span}(1)^\perp$; see for instance \cite[Def. 4.1.9 \& Thm 4.1.10]{horn2013matrix}. We note that the vectors $x$ and $y$ cannot be constant, because then $\hat{x}=0$ results in a division by zero.

To prove the bounds when $W$ is positive definite on $\operatorname{span}(1)^\perp$, we invoke the Cauchy-Schwarz inequality
$$
\left(\sum W_{ij} \hat{x}_i\hat{y}_j\right)^2 \leq \left(\sum W_{ij}\hat{x}_i\hat{x}_j\right)\left(\sum W_{ij}\hat{y}_i\hat{y}_j\right).
$$
Taking the square root, we find that $-\sqrt{\sigma_{x,W}\sigma_{y,W}}\leq \sum W_{ij}\hat{x}_i\hat{y}_j\leq \sqrt{\sigma_{x,W}\sigma_{y,W}}$ and thus $\rho_{x,y,G}\in[-1,1]$ as required. This completes the proof.\hfill$\square$

To summarize, the proposition says that once we choose formula \eqref{eq:rho-net} then it is necessary to require that $W$ is a positive definite matrix on $\operatorname{span}(1)^\perp$ and, as a side-effect of this requirement, we furthermore get that the network Pearson correlation is bounded in $[-1,1]$, similar to the classical Pearson correlation.

For a given matrix $W$, one can determine whether it is positive definite on $\operatorname{span}(1)^\perp$ or not as follows: first, calculate the matrix $\hat{W}$ with entries $\hat{W}_{ij} = W_{ij}-\tfrac{1}{n}\sum_{k}(W_{ik}+W_{kj})+\tfrac{1}{n^2}\sum_{kl}W_{kl}$, where $n$ is the size of the matrix. Second, calculate the eigenvalues of $\hat{W}$. If $\hat{W}$ has $n-1$ strictly positive eigenvalues and $1$ zero eigenvalue, then $W$ is positive definite on $\operatorname{span}(1)^\perp$ and otherwise it is not.

\subsection{Examples}\label{sec:examples}
\begin{figure}[t!]
\centering
\includegraphics[width=.33\columnwidth]{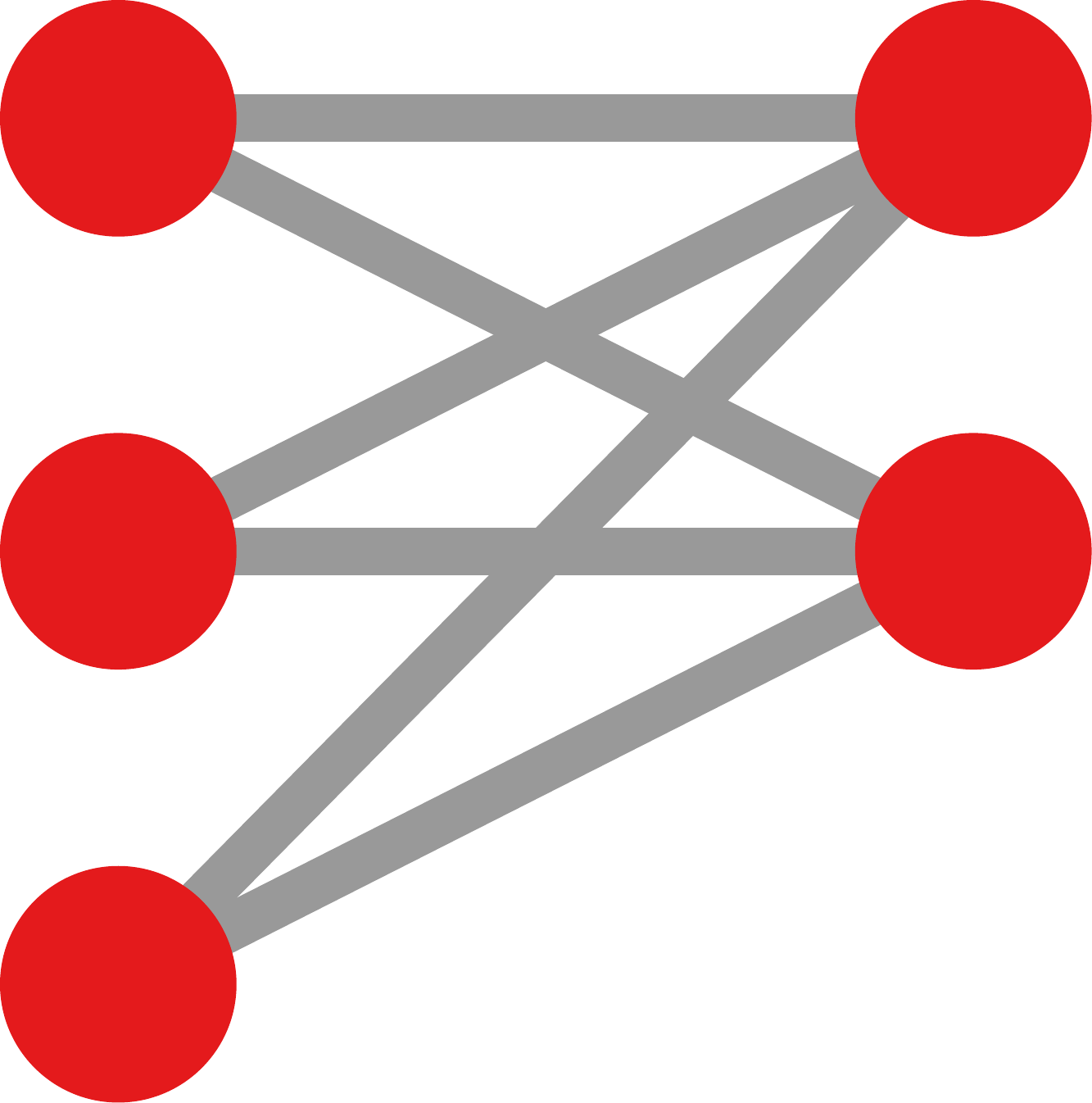}
\caption{A network with negative eigenvalues for $W = e^{-kP}$.}
\label{fig:net-example}
\end{figure}

We now give some examples where the matrix $W=e^{-kP}$ is not positive definite on $\operatorname{span}(1)^{\perp}$ and should therefore not be used to calculate network correlation coefficients. A first example is the complete bipartite graph on $2$ and $3$ nodes, shown in Figure \ref{fig:net-example}. Its adjacency matrix $A$, distance matrix $P$ and the eigenvalues of $\hat{W} = \widehat{e^{-P/4}}$ are equal to:

$$
A = \left(\begin{smallmatrix}
0 & 0 & 1 & 1 & 1\\
0 & 0 & 1 & 1 & 1\\
1 & 1 & 0 & 0 & 0 \\
1 & 1 & 0 & 0 & 0 \\
1 & 1 & 0 & 0 & 0 \\
\end{smallmatrix}\right)
\quad 
P = \left(\begin{smallmatrix}
0 & 2 & 1 & 1 & 1\\
2 & 0 & 1 & 1 & 1\\
1 & 1 & 0 & 2 & 2\\
1 & 1 & 2 & 0 & 2\\
1 & 1 & 2 & 2 & 0\\
\end{smallmatrix}\right)
\quad
\lambda\left(\hat{W}\right) = ( 0.3935,~ 0.3935,~ 0.3935,0,\textbf{-0.2}).
$$

The negative eigenvalue shows that $W$ is not positive semi-definite and thus by Proposition \ref{prop:definite} not a good weight matrix to measure the network Pearson correlation.

Examples are not limited to small and relatively simple graphs. A second and more practical example comes from a  Twitter network that records debates in the US over some political issues -- this is data from \cite{hohmann2023quantifying}. The network has $|V| = 4,458$ nodes and $|E| = 7,024$ edges, which makes it far from trivial. We calculate the eigenvalues of $\hat{W} = \widehat{e^{-P}}$ as described above and verify that we obtain a few negative ones. For instance, the minimum eigenvalue is $-5.46$. Thus, if we were to set $x$ to be the eigenvector corresponding to this minimum eigenvalue, then $\sum W \times (x \otimes x)$ would be equal to $-5.46$, and give rise to a negative variance and an imaginary correlation value.

\section{Solution}
As outlined in the introduction, one way to obtain a good weight matrix $W=e^{-kP}$ is to use a negative type metric for $P$. We first prove this formally and then describe two natural choices of negative type metrics on networks. These give rise to good weight matrices to measure network Pearson correlations.

\subsection{Negative Type Metric Spaces}
The proposal in \cite{coscia2021pearson} for a weighting matrix was $W=e^{-kP}$ where $P$ is the shortest-path distance matrix of the network. While this choice will generally not result in a well-defined correlation measure, it turns out that other types of distance matrices can work. More precisely, this is the case for so-called \emph{negative type metrics}. A negative type metric (or distance) is defined as a metric whose distance matrix $D$ is negative semidefinite on $\operatorname{span}(1)^\perp$; in other words, $-\hat{D}$ has no strictly negative eigenvalues. These metrics were introduced by Schoenberg in the context of distance geometry \cite{schoenberg1938metric} and they have many nice properties \cite{Lyons2013distance, meckes2013positive}. Several examples of negative type metrics are listen in \cite[Thm. 3.6]{meckes2013positive}.

Negative type metrics give rise to positive definite matrices as follows:
\begin{proposition}[{\cite[Thm 3.3]{meckes2013positive}}]\label{prop:negative}
The matrix $e^{-kD}$ (with entrywise exponential) is positive definite for all $k>0$ if $D$ is the distance matrix of a metric space of negative type.
\end{proposition}
If a matrix is positive definite, it is also positive definite on $\operatorname{span}(1)^T$. This means that weight matrices $e^{-kD}$ for negative type metrics $D$ can be used to measure network correlations. We also remark that the statement is even stronger: the entrywise exponent $e^{-kD}$ of a distance matrix is positive definite for all $k>0$ \textit{if and only if} if the distance has negative type. Thus, if we choose the exponential of a distance matrix as a weighting matrix, then negative type metrics are the only metrics that lead to a well-defined correlation.

As noted before, we can check whether a distance matrix has negative type by calculating the matrix $-\hat{D}$ and checking whether it has any negative eigenvalues. In what follows, we mention two examples of negative type metrics which are relevant in the context of networks.

\subsection{Effective Resistance}
A first option to obtain a good weight matrix $W$ is to use the effective resistance distance \cite{klein1993resistance, devriendt2022effective}. The effective resistance between two nodes is the resistance of the total system when a unit voltage is connected across the nodes, where the graph $G$ is interpreted as an electrical circuit with 1 Ohm resistors as edges (in the case of unweighted undirected graphs) \cite{ellens2011effective}.

The effective resistance is a negative type metric. This was shown for instance by Fiedler in \cite{fiedler2011matrices} and was discussed in detail in \cite{devriendt2022graph}. It follows from Proposition \ref{prop:negative} that if $\Omega$ is our matrix recording all pairwise effective resistance values, then $W = e^{-k\Omega}$ is positive definite and thus a valid weight matrix.

\subsection{Node Embeddings}
A second example of negative type metrics are the distances between points in Euclidean space $\mathbb{R}^d$; this example is included in the list of negative type metric spaces in \cite[Thm. 3.6]{meckes2013positive} and a number of proofs are summarized in \cite{Lyons2013distance}. This case of negative type metric spaces can be relevant in the context of networks when an embedding of the network is given. In many of today's machine learning pipelines for analyzing networks and related data, one of the first steps includes an \emph{embedding} of the nodes of the network in Euclidean space (or some other spaces \cite{bronstein2017geometric}). The goal of this embedding is that nodes which are ``close together'' or ``similar'' in the network also end up close together in the embedding space. After this embedding, the network data can then be analysed in the much familiar setting of vectors in Euclidean space, where many standard techniques are available.

It then follows from Proposition \ref{prop:negative} that, if $\Delta$ is our matrix recording all pairwise Euclidean distances between embedded nodes, then $W = e^{-k\Delta}$ is positive definite. Depending on the specific algorithm used to generate the node embeddings, one would get a different (positive) estimate for $\sum W \times (x \otimes x)$ for the bipartite $2 \times 3$ graph and the network in Figure \ref{fig:net-example}.

Note that many algorithms to derive node embeddings -- for instance DeepWalk \cite{perozzi2014deepwalk} -- are not deterministic, because they are based on random walks. If one needs a deterministic result, they can use effective resistance. In fact, effective resistances can also be interpreted as coming from Euclidean embeddings; this is often referred to as the commute time embedding \cite{Qui_2007_clustering}. However, as we pointed out above, one might already have calculated the embeddings for other purposes. If that is the case, then they would obtain a correlation measure ``for free'' without having to perform the computationally intensive effective resistance calculation. So the two approaches we sketch here -- effective resistance and node embeddings -- are both suitable for different scenarios.

\section{Conclusion}
In this note we have shown that, if one wants to calculate the network correlation proposed in \cite{coscia2021pearson}, they need to choose a positive definite $W$ matrix when representing the graph's topology. The original suggestion of using the matrix of pairwise shortest path distances is invalid because such choice leads to a $W$ that is not positive definite. We show that using pairwise effective resistance or Euclidean distances between node embeddings are valid alternatives.

\section*{Acknowledgements}
We thank Andrea Luppi for pointing the potential issue with negative variance values, and for the fruitful discussion that followed.

\bibliographystyle{plain}

\end{document}